\documentclass[aps,pre,twocolumn,groupedaddress]{revtex4}
\usepackage{color}
\def\aa{\AA\,} 
\def\eR{\textsf{R}\,}
\newcommand{\be}{\begin{equation}} 
\newcommand{\ee}{\end{equation}} 
\newcommand{\bea}{\begin{eqnarray}} 
\newcommand{\eea}{\end{eqnarray}}
\usepackage{graphicx} 
\begin{document}
\def\Journal#1#2#3#4{{#1} {\bf #2,} { #3} (#4).} 
 
\def\BiJ{ Biophys. J.}                 
\def\Bios{ Biosensors and Bioelectronics} 
\def\LNC{ Lett. Nuovo Cimento} 
\def\JCP{ J. Chem. Phys.} 
\def\JAP{ J. Appl. Phys.} 
\def\JMB{ J. Mol. Biol.} 
\def\CMP{ Comm. Math. Phys.} 
\def\LMP{ Lett. Math. Phys.} 
\def\NLE{{ Nature Lett.}} 
\def\NPB{{ Nucl. Phys.} B} 
\def\PLA{{ Phys. Lett.}  A} 
\def\PLB{{ Phys. Lett.}  B} 
\def\PRL{ Phys. Rev. Lett.} 
\def\PRA{{ Phys. Rev.} A} 
\def\PRE{{ Phys. Rev.} E} 
\def\PRB{{ Phys. Rev.} B} 
\def\PD{{ Physica} D} 
\def\ZPC{{ Z. Phys.} C} 
\def\RMP{ Rev. Mod. Phys.} 
\def\EPJD{{ Eur. Phys. J.} D} 
\def\SAB{ Sens. Act. B} 
\title{Topological change and impedance spectrum of rat olfactory {{receptor}} I7:
A comparative analysis with bovine rhodopsin and bacteriorhodopsin}
\author{Eleonora Alfinito, Cecilia Pennetta and Lino Reggiani}
\affiliation{Dipartimento di Ingegneria dell'Innovazione, Universit\`a del
Salento, 73100 Lecce, {Italy, EU}\\ CNISM - Consorzio Nazionale Interuniversitario per le Scienze Fisiche della Materia, {via
della Vasca Navale 84, 00146 Roma, Italy, EU}}




%
\date{Received: date / Revised version: date}
%
%
\begin{abstract}
We present a theoretical investigation on possible selection of olfactory receptors (ORs) as sensing components of nanobiosensors. 
Accordingly, we generate the impedance spectra of the rat OR I7 in the native and activated state and analyze their
differences. 
In this way, we connect the protein morphological transformation, caused by the sensing action, with its change of electrical impedance.
The results are compared with those obtained by studying the best known protein
of the GPCR family: bovine rhodopsin. 
Our investigations indicate that a change in morphology  goes with a change in impedance spectrum mostly associated with a decrease of the static impedance up to about 60 \% of the initial value, 
in qualitative agreement with existing experiments on rat OR I7.
The predictiveness of the model is tested successfully for the case of recent experiments on bacteriorhodopsin.
The present results point to a promising development of a new class of nanobiosensors based on the electrical properties
of GPCR and other sensing proteins. 
\end{abstract}
%
\maketitle
\section{Introduction}
\label{intro}

Nanobiosensors constitute a new frontier in the field of electronic
devices, both for their reduced dimensions and for their high specialization
in sensing action \cite{Dinh}. 
One of the most promising challenge is producing more efficient devices
owing to a direct use of biological sensing materials such as proteins.
Sensing proteins are able to catch a specific molecule (ligand), e.g., odour, hormone,
drug, light (photons), etc., then undergoing a change of their conformation.
{\it In vivo}, the conformational change activates a chain of biochemical events
that culminates in the transmission and detection of the captured information.
{\it In vitro}, it is not easy to reproduce the entire chain and, probably, it
is also not convenient. 
Thus, the capture of a specific ligand should be monitored looking only at some 
peculiar aspects of the capture mechanism, like the conformational change in the
protein. 
\par
In previous works \cite{Nano,Kumar} we described the possibility to monitor a conformational
change by using the change of the impedance spectrum associated with the single protein. 
Our model was based on the conjecture that if the protein is able to sustain a charge transport, then
the macroscopic electrical response, specifically the impedance spectrum, would depend
on the path followed by the charges and thus on the protein structure. 
When the structure changes, the impedance spectrum is expected to change in a correlated way \cite{Nano}.

\par
In the present paper we apply the model to a couple of sensing proteins,
the I7 rat olfactory receptor \cite{Gaillard} and the bovine rhodopsin \cite{Santosh}. 
Both these proteins
pertain to the huge family of the seven-helices transmembrane receptors,
the so called G protein coupled receptors (GPCRs) \cite{Lefk2000}. 
Therefore, they share a similar behavior in the sensing action:
Both proteins undergo a conformational change when going from the native to the activated state, then activating a G protein;
the former when capturing a specific odorant, the latter when capturing a photon.
As a consequence, it is potentially possible to reveal their conformational
change by measuring some variations of their electrical properties.
In this respect, it is crucial to determine whether this change is sufficiently large to be detectable within
the experimental resolution, thus making possible to transduce the sensing action of the protein into an electrical signal.
Recent experiments \cite{Hou07} have related 
the sensing action of the rat OR I7 { to the capture of
a specific ligand. The OR I7 was deposited on a self-assembled 
multilayer (SAM) and the variation of its impedance spectrum was revealed
by means of an electrochemical impedance spectroscopy (EIS) characterization}.
These results disclose the possibility of using proteins as very refined sensors, able to determine, through an electrical signal, the presence and the concentration of the substance to be detected.
\par
The paper is organized as follows.
Section II briefly recalls the theoretical model.
Sections III and IV report the results together with their physical interpretation.
Major conclusions are drawn in Sect. V.
\section{Theory}
\label{sec:1}
Our aim is to determine the variation of the protein impedance spectrum \cite{Barsoukov}
subsequent to a conformational change, by taking into account only geometrical changes in the tertiary
structure of the protein. 
To this purpose, we assume that the 3D structures of the protein in its native and activated state
are known from the protein data bank (PDB) available in literature \cite{Berman} or similarly.
In the present case, since the 3D structures are only partially known {both
for rhodopsin and OR I7}, we made use of a homology modeling for both the proteins \cite{Sali},
as detailed later.
Then we introduce a two step procedure.
\par
First, the protein in an assigned state is represented by means of a topological network (graph). 
Each node of this graph corresponds to a single amino acid:
 If two amino acids are closer than an assigned cut-off distance, {\eR}, then a link is drawn 
between the corresponding nodes.
The distance {\eR} definitely selects the connectivity of the network, and 
is here taken as an adjustable parameter of the model.  
Accordingly, the choice of its value has to be headed by some physical constraints
and finally fixed by comparison with experiments. 
At this stage, we limit ourselves to observe that the network is built
up with a strategy analog to that of random graphs, by using 
{\eR}, instead of a constant probability, to select the effective links \cite{Albert}. 
Therefore, for this model, we have the possibility to recover  some of the results obtained in the 
context of random and complex networks \cite{Albert}.
\par
Second, by associating an elemental impedance with each link, the graph becomes an impedance  network. 
In other words, each link mimics a privileged channel for the  charge transfer and/or charge 
polarization \cite{Churg, Kumar}.
By taking appropriate contacts and applying an external bias, the electrical network is solved 
within a linear circuit analysis. 
\par
In a simple approach, the spatial position of each node is taken to
correspond to the $\alpha$-carbon of each amino acid. 
Each link connecting two nodes is replaced by an elementary impedance consisting of  an Ohmic resistance connected 
in parallel with a parallel plate capacitor, filled with a dielectric. {In such a way, we can describe both charge motion and electrical polarization.}
The elementary impedance between nodes $i$ and $j$,  $Z_{i,j}$, 
is given explicitly by \cite{Kumar}:
\be
Z_{i,j}={l_{i,j}\over {\mathcal{A}}_{i,j}}   
{1\over (\rho^{-1} + i \epsilon_{i,j}\, \epsilon_0\omega)}  
\label{eq:1}
\ee              
where ${\mathcal{A}}_{i,j}=\pi ({\eR}^2 -l_{i,j}^2/4)$, is the cross-sectional
area between two spheres of radius {\eR} centered on the $i$-th and $j$-th node, respectively;
$l_{i,j}$ is the distance between these centers, $\rho$ is the 
resistivity, taken to be the same for every 
amino acid, with the indicative value of 
$\rho = 10^{10} \ \Omega$ m; $i=\sqrt{-1}$ is the imaginary unit,  
$\epsilon_0$ is the vacuum permittivity,
and  $\omega$ is the circular frequency of the applied voltage. 
The relative dielectric constant of the couple of $i,j$ amino 
acids, $\epsilon_{i,j}$, is expressed in 
terms of the intrinsic  polarizability of the $i,j$ amino acids \cite{Song}.
\par
By positioning the input and output electrical contacts, respectively, on the first and last node
(corresponding to the first and last amino acid of the protein sequential structure), 
the network is solved within a linear Kirchhoff scheme and its global impedance spectrum
is calculated within the convenient frequency range 0-100 kHz.
For a given interaction radius, the spectrum depends on the network structure. 
The change of the protein state, by implying a change in the network structure, will lead to a different spectrum of the global impedance.
Such a variation of the impedance spectrum is here taken as an estimate of the electrical response consequent the
protein conformational change.
Since the resistivity associated with an amino acid is taken to be independent of geometry and applied
voltage, the above estimate should be considered as a reference threshold value.
By considering more detailed microscopic mechanisms of charge transport \cite{Warshel}, like tunneling, 
and/or the possible influence of the medium surrounding the protein, the 
electrical sensitivity to the conformational change can be further modulated and, in particular, amplified.
The role of the medium (lipids) surrounding the protein is here neglected.
Indeed, the medium is not involved in the conformational change and its effect cancels when considering
the variation of the impedance as done here. 
\begin{figure}
\resizebox{0.45\textwidth}{!}{%
  \includegraphics{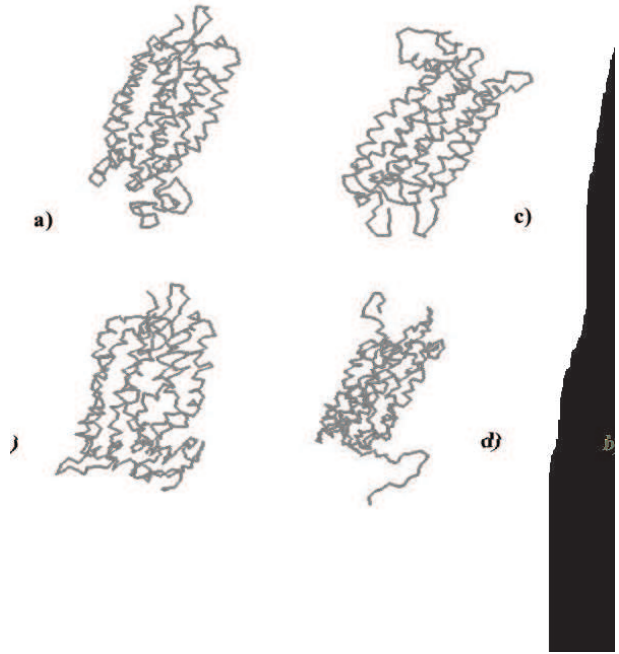}
}
\caption{Sketch of the backbone of bovine rhodopsin and rat OR I7 obtained with the homology modeling.
 a), b) refer to bovine rhodopsin  in the native and activated states, respectively; c), d) refer
to  rat OR I7 in the native and activated states, respectively.}
\label{fig:1}       
\end{figure}

\subsection{Materials and methods}
\label{sec:1.2}
The primary structure, i.e., the sequence of amino acids, is  presently available for many proteins.
However, the tertiary structure, i.e., the spatial organization, is known only for a very few of them and often is rather incomplete \cite{Berman}.  
In particular, when the representations of different states of 
the protein are given, they mostly refer to different experimental conditions ({X-ray resolution, pH,} temperature, etc).
The body of all these factors produces a large error bar in the assessment of the protein topology and in 
the possibility of faithfully discriminate different configurations \cite{Nano}. \begin{figure}
\resizebox{0.45\textwidth}{!}{%
  \includegraphics{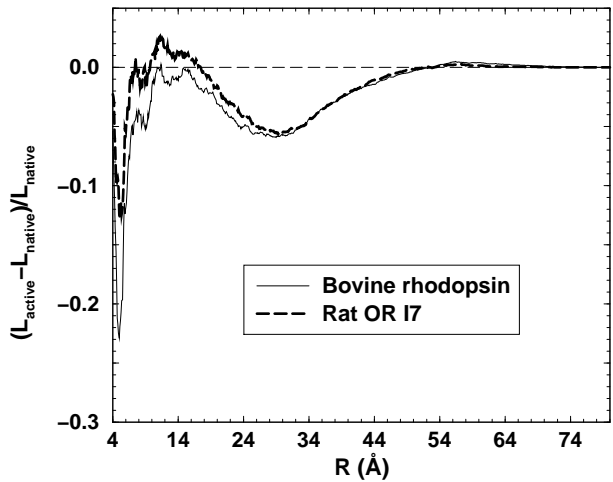}
}
\caption{Variation of the link total number vs the interaction radius for bovine rhodopsin and rat OR I7. 
The horizontal dashed line serves as a reference for negligible variations.}
\label{fig:2}       
\end{figure}
At present there are not complete representations of the tertiary structure
for both the native and activated state of bovine rhodopsin, taken in the
same experimental conditions. 
Even worse is the state of art of rat OR I7, for which
only the primary sequence is taken by experiments. 
At present, for both the proteins, only 
the tertiary structure of the native states has been predicted, with first principle methods \cite{Vaidehi}.
For a reliable modeling, we require that both the native and activated 3D structures originate by the same experimental or theoretical conditions. 
For this purpose, the tertiary structures of native and activated states for rhodopsin
and OR I7 were obtained by means of the {MODELLER} software \cite{Sali}. 
The template PDB entries were the same for both the proteins: 1JFP and 1LN6 for the native and activated state, respectively, complemented with 1F88, chain A. 
Both 1JFP and 1LN6 were produced in the same experimental conditions.

\par
Figures 1 (a)-(d) report the backbone of bovine rhodopsin and rat OR I7 in their natural and activated states,
as determined by the homology modeling.
\par
The method we adopt for comparing the change in impedance spectrum associated with the change of configuration 
is the analysis in the corresponding Nyquist plot. 
This plot is obtained by drawing the negative imaginary part versus the real part of the global impedance, 
within a given frequency range (typically from 1 mHz to 100 kHz as in experiments
\cite{Hou06,Hou07}).
For a single \textit{RC} parallel circuit, the shape of the Nyquist plot appears as a perfect semi-circle. 
In the present case, owing to the presence of many different \textit{RC} circuits, the plots
slightly depart from the perfect semi-circle shape, as more generally described in terms
of  Cole-Cole plots \cite{Nano, Cole}.
\section{Results and Discussion}
\label{sec:2}
The topological network obtained from the protein mapping is in general dependent on the value of {\eR}. 
Indeed, by increasing {\eR}, the degree of each node, i.e., the number of links spanned by each node \cite{Albert}, increases in a way that reflects the node distribution in the network. For very large values of {\eR}, the node
degree saturates to the expected value $N-1$, with $N$ the number of nodes.
Consequently, the total link number shows a characteristic
{sigmoid-like} profile, which saturates to the value $N(N-1)/2$. 
\par
Figure 2 reports the relative variation  of the total link number associated with the native and the 
activated state as function of  {\eR}, for rat OR I7 and rhodopsin, respectively.
For small values of {\eR}, e.g., less than 6 \aa, 
the network is poorly connected, which implies a  large sensitivity of the number of links on {\eR}. 
On the other hand, for large values of {\eR}, e.g., larger than about 20 \aa, 
most of the distances between nodes are smaller than {\eR} and quite all the nodes are connected: {It} implies
a smooth variation of the total number of links. 
At these asymptotic values of {\eR},  differences in the relative distance between amino acids, induced 
by a conformational change, cannot be appreciated by the model. 
In the intermediate region of {\eR} values, the number of links strictly depends on the distance between amino acids
and thus the model makes  possible to resolve the different states of the protein. 
Accordingly, the values of {\eR} useful for making the network able to capture the
variation of impedance subsequent to a conformational change, should be chosen in 
the intermediate region 6-20 \aa, which also overlaps with the range of 
physical interactions among amino acids \cite{Tirion,Nano,Juanico}.
\par
Figure 3 reports the degree distribution for relevant values of 
the interaction radius.
\begin{figure}
\resizebox{0.45\textwidth}{!}{%
  \includegraphics{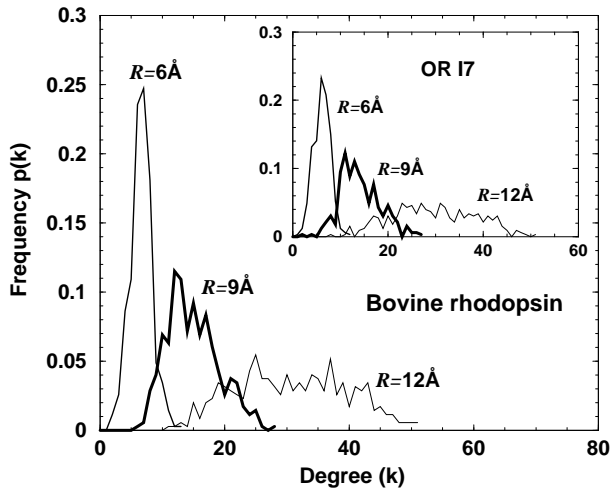}
}
\caption{Degree distribution for the native states of bovine rhodopsin and rat OR I7 (in the insert). 
The curves correspond to  values of the interaction radius in the range from 6 to 12 \aa.
Distributions obtained with larger values of {\eR} loose of resolution in this scale.}
\label{fig:2}       
\end{figure}
Here we observe that, for small values of {\eR}, say 6 \aa, the distribution is rather peaked
implying that each amino acid interacts practically with the same number of nearest neighbors, independently 
of its position in the protein.
The distribution broads at increasing values of {\eR}, say 9-12  \aa, where, in any case,
several small but detectable peaks appear in the distribution profile.
These peaks signal the different clustering of the protein main structures (helices and sheets).
Accordingly, for the values of {\eR} in the range 6-12 \aa, the network is maximally able to monitor the 
internal structures of the protein.
When {\eR} is further increased, e.g., {\eR} larger than 12 \aa,  
the degree distribution becomes very broadened and no longer able to provide information on 
the internal structures of the protein. 
\begin{figure}
\resizebox{0.45\textwidth}{!}{%
  \includegraphics{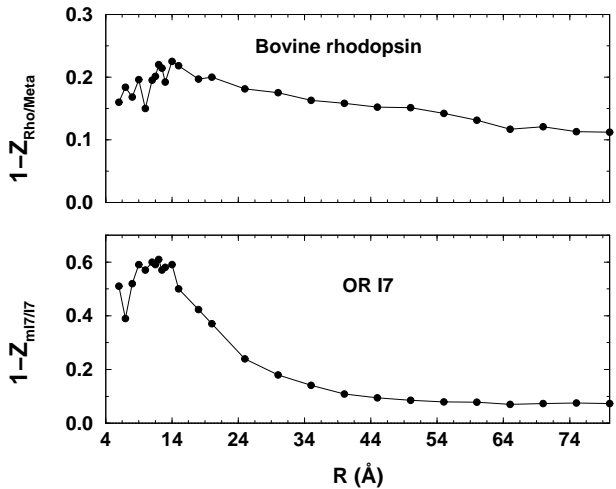}
}
\caption{Relative variation of the static impedance of bovine rhodopsin and rat OR I7 vs the interaction radius}
\label{fig:5}       
\end{figure}
\begin{figure}
\resizebox{0.45\textwidth}{!}{%
  \includegraphics{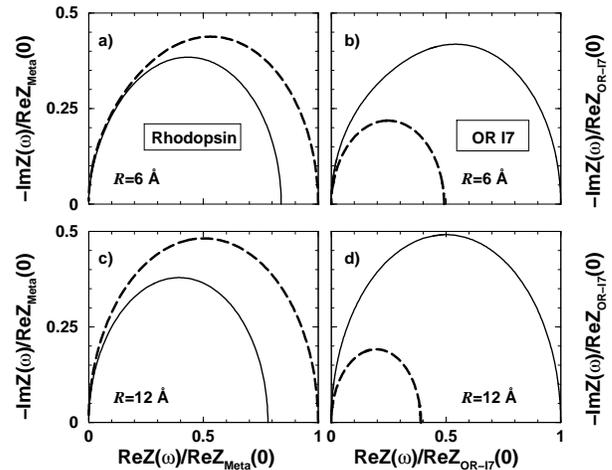}
}
\caption{Nyquist plots for the rhodopsin [frames (a) and (c)] and rat OR I7 [frames (b) and (d)].
Both the real and imaginary parts of impedance have been normalized to the value of the static impedance of the 
corresponding activated state of bovine rhodopsin and rat OR I7, respectively. 
Continuous lines refer to the native state, and dashed
lines to the activated state, respectively.}
\label{fig:4}       
\end{figure}
{Finally, in the
case the conformational change mainly involves 
displacements of nearest neighbors,  the topological network has to be constructed by choosing values of \eR in the range 6-9 \aa. This choice leads to quite different patterns for the natural and activated structure.} 
Otherwise, when the conformational change involves displacements of entire structures 
like helices, the values of the interaction radius should be chosen 
in a slightly wider range, say 9-12 \aa. 
\par
Figure 4 reports the relative variation of the static impedance (resistance)
as function of the interacting radius for rhodopsin and rat OR I7. 
We can observe that the maximal variation is in the
range 5-20 \aa, for both the proteins. 
This is in accordance with what shown in Figs. 2 and 3. 
Furthermore, from this figure it emerges that the variation of resistance 
is never smaller than 10 $\%$ for both the proteins and, as a consequence, their conformational
change is large enough to be revealed from variations of their electrical properties, 
in particular, from EIS measurements.
We notice that the variation of resistance is significantly larger than the link difference,
indicating the greater relevance of the specificity (length and position) of the links rather than of their number. 
This outcome downsizes the importance  of the precise value of {\eR}:
In fact, the electrical resolution of the conformational change is practicable within a wide range of {\eR} values. 
Therefore, the qualitative accordance with experimental data is preserved to a large extent, while the quantitative agreement can be achieved by fine tuning the value of {\eR}.
\par
On this basis, to evidence the most relevant features that a conformational change in the protein 
can induce on its impedance  spectrum, we take two {\eR} values,  6 \aa and 12 \aa for both rhodopsin and rat OR I7.
Figure 5 compares the Nyquist plots for the native and activated states of rhodopsin [Figs. 5 (a) and (c) in the left column] and of rat OR I7 [Figs. 5 (b) and (d) in the right column].
For both proteins the shape of the plots is close to a semicircle, apart a small squeezing for {\eR} = 6 \aa.
This result indicates that a smaller distribution of \textit{RC} circuits is reflected in a deviation from the otherwise
perfect  semicircle shape, as expected \cite{Nano}.
An increasing resolution with increasing values of {\eR} is evidenced for both proteins, with the best value obtained for {\eR} = 12 \aa : This result confirms the presence of a large conformational change \cite{Nano,Cole}
In particular, for rat OR I7 the resistance between the configuration is found to exhibit a significative larger difference than for rhodopsin at both the considered values of {\eR}.
Finally, we can notice that while in rhodopsin the native state shows a smaller
value of the static impedance with respect to the activated state, in OR I7 it
happens the opposite. 
The origin of this difference can be traced back to the shrinking trend
of the activated structure of OR I7, and in the opposite behavior
of rhodopsin,  as can be guessed from Fig. 1. 
The theoretical results for rat OR I7 compare well with recent EIS experiments carried out on OR I7 samples 
deposited on SAM in the presence of specific odorants, heptanal and octanal
\cite{Hou07}. 
The experimental Nyquist spectra exhibited a near semicircle shape in the region of intermediate low frequencies with a decrease of the resistance for about 20 \% in the presence of octanal at a concentration of 10$^{-4}$ {mol/l}, in qualitative agreement with the expectation of the present model. 
The experiments gave evidence also of a Warburg resistance at the smallest frequency, here neglected. 
At this stage, since we are comparing macroscopic measurements with a microscopic model (single protein) the 
qualitative agreement found is considered to be satisfactory by confirming,
in essence, that {\it the capture of an odorant by an OR belonging to the GPCR family can be monitored by means of electrical measurements}.

\subsection{Global properties}
\label{sec:2.4}
We conclude this section by reporting the  behavior of global protein conductance and capacitance, 
as functions of {\eR}.
Figure 6 shows the static conductance, $G=1/Z(0)$, versus the reduced cut-off radius ({\eR}-{\eR}$_{0}$), both for the 
native and the activated state of rhodopsin, where {\eR} less than {\eR}$_{0} \approx$3.6 \aa is the value
of {\eR}, which corresponds to zero conductance (i.e., disconnected network). 
The static conductance is found to follow a power law behavior: $G \propto ({\eR}-{\eR}_0)^{\gamma}$, 
where the exponent $\gamma$ takes two different values for  two different regions of {\eR} values: 
$\gamma\,\simeq \,2$ for $(\eR-\eR_0) >$ 30 \aa , and  $\gamma\,\simeq \,3$ for $({\eR}-{\eR}_0) <$ 20 \aa 
(see Fig. 6), for both the native and activated states.
We can explain this result by observing that from Eq. (\ref{eq:1}) it emerges a ${\eR}^{2}$-dependence of the single conductance. 
Then, for increasing number of links, the parallel connections increase more than those in series  and this produces a 
further ${\eR}$-dependence.  
When the number of links  approaches  the saturation value, only the  ${\eR}^{2}$-dependence survives.
As expected, the global capacitance is found to follow the same behavior of the conductance, as shown in Fig. 7.
Analogous results have been obtained, but not shown, for rat OR I7. 
\begin{figure}
\resizebox{0.48\textwidth}{!}{%
  \includegraphics{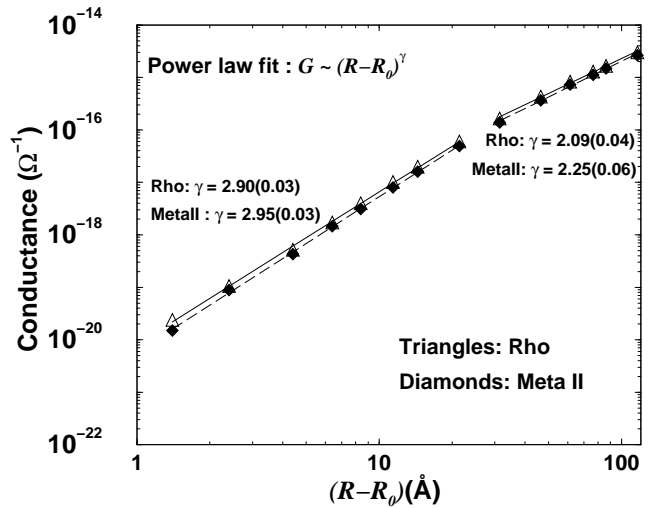}
}
\caption{Static conductance of the bovine rhodopsin protein. 
The symbols refer to calculations; the straight lines are the best fit curves. 
The continuous line refers to the native rhodopsin (Rho) and the dashed line refer to the activated rhodopsin (Meta II).}
\label{fig:6}       
\end{figure}
\begin{figure}
\resizebox{0.48\textwidth}{!}{%
  \includegraphics{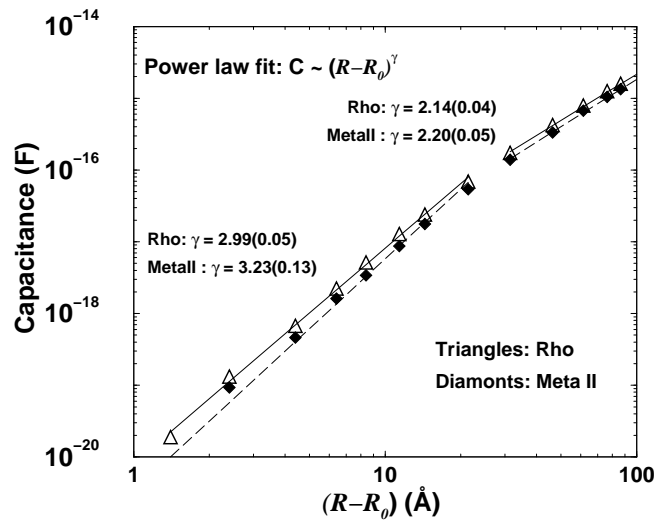}
}
\caption{Global capacitance of the bovine rhodopsin protein. 
The symbols refer to calculation; the straight lines are the best fit curves. 
The continuous line refers to the native rhodopsin (Rho) and the dashed line refer to the activated rhodopsin (Meta II).}
\label{fig:7}       
\end{figure}
\section{A further test of the network model}
\label{3}
As a further validation of the network model, we have considered the case of bacteriorhodopsin.
The change of its electrical properties due to the conformational change induced by green light 
has been recently investigated through current voltage measurements \cite{Jin,
Gomila}. 
Bacteriorhodopsin, i.e., a seven helices light receptor like rhodopsin, is present
in {archaea} instead of mammals and works with a different mechanism of signal transmission. 
Like rhodopsin, its light-sensitive part is the retinal, which is located deep inside the protein. 
When retinal captures photons, it changes its shape and induces the protein
conformational change. 
However, in bacteriorhodopsin retinal changes its shape from straight to bent, just the opposite of
what happens in rhodopsin \cite{Berman}. 
Accordingly, for bacteriorhodopsin we expect a  pattern of the Nyquist plot associated with the transformation, 
opposite to that exhibited by bovine rhodopsin, see Fig. 5, where the activated state exhibits a 
static impedance greater than that of the native state. 
To this purpose, Fig. 8 reports the Nyquist plot for
two different couples of native and activated representations of bacteriorhodopsin, 1FBB (native) - 1FBK (acti\-vated) 
and 2NTU (native) - 2NTW (activated), for {\eR} = 6 \aa. 
The plots confirm that the static impedance of the native state is significantly greater (up to 15 \%)
than that of the
activated state, which is, in fact, a behavior opposite to that of bovine rhodopsin.
On the other hand, this result conforms with the expectations of the topological analysis
and agrees with the experimental evidence of an increase of conductance when passing from the native (i.e,. in {the dark}) to
the activated state (i.e., in the presence of green light) \cite{Jin}.
These findings 
are taken as a further validation of the network model here reported.
\begin{figure}
\resizebox{0.45\textwidth}{!}{%
  \includegraphics{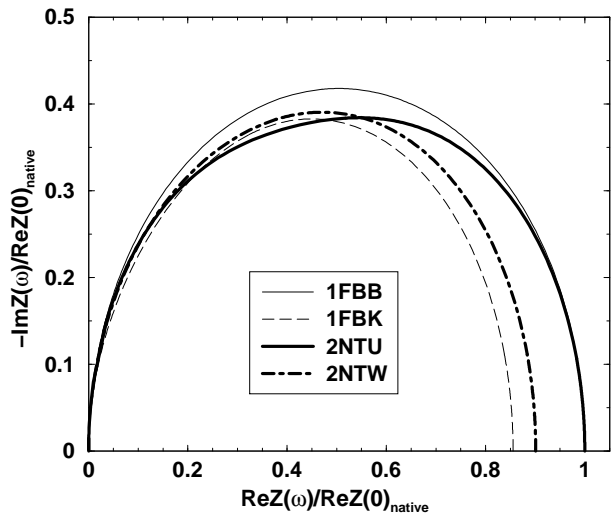}
}
\caption{Nyquist plots for two couples of native and activated states of bacteriorhodopsin:
1FBB - 1FBK  and 2NTU - 2NTW, native and activated states of the protein, respectively.
 Plots have been obtained with {\eR}=6 \aa.}
\label{fig:8}       
\end{figure}
\section{Conclusions}
We have investigate the impedance spectrum of rat {OR I7}
and carried out a comparative analysis
with that of the homologous GPCR, bovine rhodopsin. 
The {used model is a powerful tool} in the analysis of the conformational changes these proteins undergo 
when exerting their function. In fact, the model allows an estimation of the magnitude of electrical 
impedance variation associated with the activation of the protein. 
These kind of measurements, at present
quite difficult to be carried out,  are, anyway, of great interest. 
In fact, they can disclose the possibility to produce a new generation
of sensors of nanometric dimensions and sensible to very small quantities of hormones, neurotransmitters, toxic products, and so on \cite{Kumar}.

When applied to rhodopsin and rat OR I7, the impedance network predicts 
the possibility to resolve the different configurations pertaining to the native and activated state,
by the difference exhibited by the corresponding Nyquist plots.
The resolution level is predicted to achieve a maximum of about 60 \% in the variation of
the static impedance.
EIS of rat OR I7 sensing action has
 evidenced changes of the impedance 
spectrum up to about 20 \% \cite{Hou07} when specific odorants have been introduced on the samples 
anchoring the ORs. 
These results are taken as an indirect validation of the present modeling.
{As a further validation test, the methodology here proposed was also applied to the bacteriorhodopsin. The model correctly predicts the opposite behavior of the static impedance of this protein compared to that of bovine rhodopsin.} 
Also this behavior is in qualitative agreement with experimental measurements \cite{Jin}.
As final remark, we mention the possible role played by the surrounding media of the protein in determining the absolute value of the impedance.
Here we neglected this effect, which we expect to be rather insensitive to the conformational change.
\begin{acknowledgments}
The authors acknowledge Dr. Edith Pajot-Augy and Dr. Gabriel Gomila for useful discussions on the subject
and  Dr. Vladimir Akimov for modeling the tertiary structures of rat OR I7
and bovine rhodopsin.

This research was partially supported by the MIUR PRIN ``Strumentazione 
elettronica integrata per lo studio di  variazioni conformazionali di proteine tramite
misure elettriche'' Prot. No.~2005091492. 
\end{acknowledgments}

%

%

\end{document}